\documentclass[aps,showpacs,nofootinbib,superscriptaddress]{revtex4}
\usepackage[dvips]{graphicx}
\usepackage{dcolumn}
\usepackage{multirow}

\begin{document}

\title{ Heavy quark symmetry constraints on semileptonic form factors
and decay widths of doubly heavy baryons } \author{ E. Hern\'andez}
\affiliation{Grupo de F\'\i sica Nuclear, Departamento de F\'\i sica
Fundamental e IUFFyM, Universidad de Salamanca, E-37008 Salamanca,
Spain.}  \author{J. Nieves} \affiliation{Departamento de F\'{\i}sica
At\'omica, Molecular y Nuclear, Universidad de Granada, E-18071
Granada, Spain.} \author{ J. M. Verde-Velasco} \affiliation{Grupo de
F\'\i sica Nuclear, Departamento de F\'\i sica Fundamental e IUFFyM,
Universidad de Salamanca, E-37008 Salamanca, Spain.}
\begin{abstract} 
  \rule{0ex}{3ex} We show how heavy quark symmetry constraints on
  doubly heavy baryon semileptonic decay widths can be used to test
  the validity of different quark model calculations. The large
  discrepancies in the results observed between different quark model
  approaches can be understood in terms of a severe violation of heavy
  quark spin symmetry constraints by some of those models.
\end{abstract}

\maketitle

\section{Introduction}

\begin{table}
\begin{tabular}{ccccc||ccccc}
\hline
Baryon\hspace{.3cm} & Quark content & $S_h$ & $J^\pi$ & Mass [MeV] &
 Baryon\hspace{.3cm} & Quark content & $S_h$ &
$J^\pi$& Mass [MeV]\\ \hline
$\Xi_{cc}$ & c~c~l & 1 & 1/2$^+$ &3612& $\Omega_{cc}$ & c~c~s & 1 &
1/2$^+$ &3702 \\ \hline
$\Xi_{cc}^*$ & c~c~l & 1 & 3/2$^+$ &3706& $\Omega_{cc}^*$ & c~c~s & 1
& 3/2$^+$ & 3783 \\ \hline
$\Xi_{bb}$ & b~b~l & 1 & 1/2$^+$ &10197& $\Omega_{bb}$ & b~b~s & 1 &
1/2$^+$ & 10260 \\ \hline
$\Xi_{bb}^*$ & b~b~l & 1 & 3/2$^+$ &10236& $\Omega_{bb}^*$ & b~b~s & 1
& 3/2$^+$ & 10297\\ \hline
$\Xi_{bc}$ & b~c~l & 1 & 1/2$^+$ & 6919& $\Omega_{bc}$ & b~c~s & 1 &
1/2$^+$ &6986 \\ \hline
$\Xi_{bc}'$ & b~c~l & 0 & 1/2$^+$ & 6948&$\Omega_{bc}'$ & b~c~s & 0 &
1/2$^+$  & 7009\\ \hline
$\Xi_{bc}^*$ & b~c~l & 1 & 3/2$^+$ &6986 & $\Omega_{bc}^*$ & b~c~s & 1
& 3/2$^+$ & 7046\\ \hline
\end{tabular}
\caption{Quantum numbers of doubly heavy baryons analyzed in this
study. $J^P$ is the spin parity of the baryon, and $S_{h}$ is the spin
of the heavy degrees of freedom. $l$ denotes a light $u$ or $d$
quark. Mass predictions from Ref.~\cite{conrado07} obtained  
using the AL1 interquark potential of Ref.\cite{SS94} are also given.}
\label{tab:summ}
\end{table}


In hadrons with a single heavy quark the dynamics of the light degrees
of freedom becomes independent of the heavy quark flavor and spin when
the mass of the heavy quark is made arbitrarily large. This is known
as heavy quark symmetry (HQS)~\cite{hqs1,hqs2,hqs3,hqs4}.  This
symmetry can be developed into an effective theory
(HQET)~\cite{georgi90} that allows a systematic, order by order,
evaluation of corrections in inverse powers of the heavy quark mass.
Ordinary HQS can not be applied directly to hadrons containing two
heavy quarks. There, the kinetic energy term, needed to regulate
infrared divergences, breaks heavy flavor
symmetry~\cite{thacker91}. Only the spin symmetry for each of the
heavy quark flavor is preserved.  The symmetry that survives is heavy
quark spin symmetry (HQSS), which amounts to the decoupling of the
heavy quark spins for infinite heavy quark masses. In that limit one
can consider the total spin of the two heavy quark subsystem ($S_h$)
to be well defined.  HQSS is sufficient to derive relations between
form factors for the decay of hadrons containing two heavy
quarks. That was first shown in Ref.~\cite{white91}, where the authors
adopted an approach where the two heavy quarks bind into a color
anti-triplet which appears as a pointlike color source to the light
degrees of freedom.  Applying the ``superflavor'' formalism of Georgi
and Wise~\cite{Georgi:1990ak,Savage:1990di,Carone:1990pv} allowed the
matrix elements of the heavy-flavor-changing weak current to be
evaluated between different baryon states. Semileptonic decays of the
$B_c$ meson were also studied using HQSS in Ref.~\cite{jenkins93}. The
formalism employed in \cite{jenkins93} has been recently extended to
describe semileptonic decays of $bc$ baryons to $cc$
baryons~\cite{flynn07}. The scheme presented in ~\cite{flynn07} does
not rely on the ``superflavor'' formalism and HQSS is naturally
implemented in it. In agreement with Ref.~\cite{white91}, the
authors\footnote{They find two differences with the results of
Ref.~\cite{white91}, which cannot be eliminated by redefining the
phases of the physical states. One difference was already pointed out
in~\cite{Sanchis-Lozano:1993kh}.}  of Ref.~\cite{flynn07} found
that spin symmetry for two heavy quarks enormously simplifies heavy to
heavy semileptonic baryon transitions in the heavy quark limit and
near the zero recoil point. As a result it is shown how an unique
function, called the Isgur-Wise (IW) function, describes an entire family
of decays involving doubly heavy baryons with total spin 1/2 and 3/2.
This imposes limitations to any quark model calculated form
factors. Besides, the fact that all baryon matrix elements are given
in term of just one function induces relations among different decay
widths that, to our knowledge, have not been exploited before to check
the validity of different quark model calculations.  This is the main
purpose of this letter.

In a recent work~\cite{conrado07} we have studied, within a
nonrelativistic quark model framework, static properties of doubly
heavy baryons and their semileptonic decays driven by the $b\to c$
transition at the quark level. For the semileptonic decays we limited
ourselves to spin 1/2 to spin 1/2 baryon transitions\footnote{In that
reference we missed a factor $1/\sqrt2$ that affected our results for
form factors. Decay widths were thus affected by a factor 2. An
erratum has been sent.}. While we have shown that our wave functions
have the correct limit for infinite heavy quark masses\footnote{In the
infinite heavy quark mass limit the baryon should look like a meson
composed of a light quark and a heavy diquark. }, we did not
check HQSS constraints on the form factors or decay widths. Here we
would like to extend our previous study on doubly heavy baryon $b\to
c$ semileptonic decays to include also doubly heavy spin 3/2 baryons
and test our model and others against HQSS predictions. These type of
decays have been studied in different relativistic quark model
approaches~\cite{ebert04,faessler01,guo98}, with the use of
HQET~\cite{sanchis95}, using QCD sum rules~\cite{kiselev02} and
three-point nonrelativistic QCD sum rules~\cite{onishchenko00}, or in
the framework of the operator product expansion using the inverse
heavy quark mass technique~\cite{kiselev00}.  Discrepancies between
the results obtained in different quark model are sometimes very
large.  Therefore, it is worthwhile to use HQSS relations among decay
widths to check the validity of the different calculations.

  In Table~\ref{tab:summ} we summarize the quantum numbers of the
doubly heavy baryons considered in this study\footnote{Note that the
definitions of $\Xi_{bc}$ and $\Xi'_{bc}$ are interchanged in some
references, with $\Xi_{bc}$ having $S_h=0$ and $\Xi'_{bc}$ having
$S_h=1$.  The same applies to $\Omega_{bc}$ and $\Omega'_{bc}$.  In
tables we always quote the results corresponding to our convention
(see Table~\ref{tab:summ}).}.

\section{Form factor decomposition}
\label{sec:ffd}

Hadronic matrix elements  can be parameterized in terms of form factors.
For $1/2 \to 1/2$ transitions the commonly used form factor decomposition reads
\begin{eqnarray}
\label{eq:1212}
\left\langle B'(1/2), r'\ \vec{p}^{\,\prime}\left|\,
\overline \Psi^c(0)\gamma^\mu(1-\gamma_5)\Psi^b(0)
 \right| B(1/2), r\ \vec{p}
\right\rangle& =& {\bar u}^{B'}_{r'}(\vec{p}^{\,\prime})\Big\{
\gamma^\mu\left(F_1(w)-\gamma_5 G_1(w)\right)+ v^\mu\left(F_2(w)-\gamma_5
G_2(w)\right)\nonumber\\
&&\hspace{1.5cm}+v'^\mu\left(F_3(w)-\gamma_5 G_3(w)
\right)\Big\}u^{B}_r(\vec{p}\,) \label{eq:def_ff}
\end{eqnarray}
with $\left|B(S), r\ \vec p\ \right\rangle$ representing a baryon
 state with three-momentum $\vec p$, total spin $S$, and spin third
 component $r$. The baryon states are normalized such that $\langle
 B(S), r'\ \vec{p}^{\,\prime}\, |\,B(S), r \ \vec{p} \rangle =
 (2\pi)^3 (E_B /m_B)\,\delta_{rr'}\,
 \delta^3(\vec{p}-\vec{p}^{\,\prime})$ being $E_B,\,m_B$ the baryon
 energy and mass. The $u^{B}_{r}$ are dimensionless Dirac spinors,
 normalized as ${\bar u}_{r'} u_r = \delta_{r r'}$. $v^\mu$, $v'^\mu $
 are the four velocities of the initial and final baryon. The three
 vector $F_1,\,F_2,\,F_3$, and three axial $G_1,\,G_2,\,G_3$ form
 factors are functions of the velocity transfer $\omega=v\cdot
 v^\prime$ or equivalently of the four momentum transfer ($q=p-p'$)
 square $q^2= m_{B}^2 + m_{B'}^2 - 2m_{B}m_{B'}\omega$.  In the decay
 $\omega$ [ $q^2$ ] ranges from $\omega=1$ $\left [ q^2=q^2_{\rm max}=
 (m_{B}- m_{B'})^2 \right]$, corresponding to zero recoil of the final
 baryon, to a maximum value given by $\omega=\omega_{\rm max}=
 (m_{B}^2 + m_{B'}^2-m_l^2)/(2m_{B}m_{B'})$ $\left [q^2=m_l^2
 \right]$, which depends on the transition, and where $m_l$ stands for
 the final charged lepton mass (we neglect neutrino masses).\\

For $1/2 \to 3/2$ transitions we follow Llewellyn Smith~\cite{LS72} to write
\begin{eqnarray}
\label{eq:1232}
&&\hspace{-1cm}\left\langle B'(3/2),r'\vec p\,'\right|\,\overline 
\Psi^c(0)\gamma^\mu(1-\gamma_5)\Psi^b(0)\left|B(1/2),r\,
\vec p\,\right\rangle=
~\bar{u}^{B'}_{\lambda\,r'}(\vec{p}\,')\Gamma^{\lambda\,\mu}
u^{B}_r(\vec{p}\,)
\nonumber\\
\Gamma^{\lambda\,\mu}=&&\hspace{-.35cm}
\left(\frac{C_3^V(\omega)}{m_{B}}(g^{\lambda\,\mu}q
\hspace{-.15cm}/\,
-q^\lambda\gamma^\mu)+\frac{C_4^V(\omega)}{m_{B}^2}(g^{\lambda\,\mu}qp'-q^\lambda
p'^\mu)+\frac{C_5^V(\omega)}{m_{B}^2}(g^{\lambda\,\mu}qp-q^\lambda
p^\mu)+C_6^V(\omega)g^{\lambda\,\mu}\right)\gamma_5\nonumber\\
&&\hspace{-.7cm}+\left(\frac{C_3^A(\omega)}{m_{B}}(g^{\lambda\,\mu}q
\hspace{-.15cm}/\,
-q^\lambda\gamma^\mu)+\frac{C_4^A(\omega)}{m_{B}^2}(g^{\lambda\,\mu}qp'-q^\lambda
p'^\mu)+{C_5^A(\omega)}g^{\lambda\,\mu}+\frac{C_6^A(\omega)}{m_{B}^2}
q^\lambda q^\mu\right)
\end{eqnarray}
with $p,\,p'$ the four-momenta of the initial, final baryon, and where
 we use the convention $g^{\mu\mu}=(+,-,-,-)$.
 ${u}^{B'}_{\lambda\,r'}$ is a dimensionless Rarita-Schwinger spinor
 normalized as $\bar{u}_{\lambda r'}\,u^\lambda_r=-\delta_{r r'}$.\\

 For $3/2\to 1/2$ transitions we have
\begin{eqnarray}
&&\hspace{-1cm}\left<B'(1/2),r'\vec p\,'\right| \,\overline 
\Psi^c(0)\gamma^\mu(1-\gamma_5)\Psi^b(0) \left|B(3/2),r
\vec p\,\right>~=
~\bar{u}^{B'}_{r'}(\vec{p}\,')\hat\Gamma^{\lambda\,\mu}
u^{B}_{\lambda\,r}(\vec{p}\,)\nonumber\\ 
&&\hat\Gamma^{\lambda\,\mu}=\gamma^0[\Gamma^{\lambda\,\mu}
(m_{B}\longrightarrow m_{B'},p\longleftrightarrow p',q
\longrightarrow-q)]^\dagger\gamma^0
\end{eqnarray}

Finally for $3/2\to 3/2$ transitions, we believe there are 25 vector
plus 25 axial form factors. The amount of form factors suggest a
different strategy in this case and thus we do not show the form
factor decomposition.
 
In Ref.\cite{conrado07} we presented results for $1/2\to 1/2$
transition form factors.  In a similar way one can evaluate all form
factors for $1/2 \longleftrightarrow 3/2$ transitions. It is not the
purpose of this work to present results for all individual form
factors. Instead, we would like to study  to what extend they obey the
restrictions imposed by HQSS.
\section{HQSS constraints on form factors for semileptonic doubly
  heavy baryon decay}
\label{sect:hqssff}
 We quote in what follows the results obtained in Ref.\cite{flynn07},
 using HQSS and near zero recoil, for the semileptonic $bc\to cc $
 baryon decay with the initial baryon at rest. There it was found that
 all hadronic matrix elements were given in terms of just one
 universal function ($\eta (\omega)$), known as the IW
 function.  Indeed  HQSS predicts
\begin{eqnarray}
\label{eq:hqss1}
B_{bc}\to B_{cc}\hspace{1cm} &&\hspace{1cm}\frac{1}{\sqrt2}\,\eta\ \bar
u'_{r'}(-\vec q\,)(2\gamma^\mu-\frac{4}{3}\gamma^\mu\gamma_5)u_r(\vec 0)\\
\label{eq:hqss2}
B_{bc}'\to B_{cc}\hspace{1cm} &&\hspace{1cm}\frac{1}{\sqrt2}\,
\frac{-2}{\sqrt3}\eta\ \bar
u'_{r'}(-\vec q\,)(-\gamma^\mu\gamma_5)u_r(\vec 0)\\
\label{eq:hqss3}
B_{bc}\to B^*_{cc}\hspace{1cm} &&\hspace{1cm}\frac{1}{\sqrt2}\,\frac{-2}{\sqrt3}\eta\ \bar
u'^\mu_{r'} (-\vec q\,)u_r(\vec 0)\\
\label{eq:hqss4}
B_{bc}'\to B^*_{cc}\hspace{1cm} &&\hspace{1cm}\frac{1}{\sqrt2}\,(-2)\,\eta\ \bar
u'^\mu_{r'}(-\vec q\,) u_r(\vec 0)\\
\label{eq:hqss5}
B^*_{bc}\to B_{cc}\hspace{1cm} &&\hspace{1cm}\frac{1}{\sqrt2}\,\frac{-2}{\sqrt3}\eta\ \bar
u'_{r'}(-\vec q\,) u^\mu_r(\vec 0)\\
\label{eq:hqss6}
B^*_{bc}\to B^*_{cc}\hspace{1cm} &&\hspace{1cm}\frac{1}{\sqrt2}\,(-2)\,\eta\ \bar
u'^\lambda_{r'}(-\vec q\,)(\gamma^\mu-\gamma^\mu\gamma_5) u_{\lambda r}(\vec 0)
\end{eqnarray}
where here $B$ stands for a $\Xi$ or $\Omega$ baryons. The IW function
which controls the $\Xi$ decays is different to that appearing in
$\Omega$ decays since the IW function depends on the light degrees of
freedom. The IW function $\eta$ is approximately one at zero recoil
($\eta(\omega=1)\approx 1$), as can be deduced from vector
conservation in the limit of degenerate $b$ and $c$
quarks\footnote{Note the authors of Ref.\cite{flynn07} missed a global
normalization factor $1/\sqrt2$. An erratum has been sent.}.

Similar results can be obtained for semileptonic $bb\to bc$
baryon decays, but there will have different IW functions because of
the heavy flavor symmetry (HFS) breaking in hadrons with two heavy
quarks.\\

Let us see the implications of the above relations for the form
factors calculated in quark models.

\subsection{$1/2\to1/2$ transitions}
Near zero recoil the three vector structures $\gamma^\mu$, $v^\mu$ and
$v'^\mu$ present in Eq.(\ref{eq:1212}) give, up to corrections
proportional to $|\vec q\,|$ that cancel near zero recoil, the same
contribution. On the other hand, the Dirac's structure of the axial
form factors $G_2$ and $G_3$, and due to the anti-diagonal nature of
$\gamma_5$, give contributions that are again proportional to $|\vec
q\,|$ and thus cancel near zero recoil.  To the extent that for the
actual heavy quark masses we are close enough to the infinite heavy
quark mass limit, Eqs.(\ref{eq:hqss1},\ref{eq:hqss2}) imply the
following restrictions on form factors
\begin{eqnarray}
\label{eq:ff1212}
B_{bc}\to B_{cc} \hspace{1cm} &&\hspace{1cm} 
\frac{1}{\sqrt2}(F_1+F_2+F_3)=\frac{3}{2\sqrt2}G_1=\eta\\
B'_{bc}\to B_{cc} \hspace{1cm} &&\hspace{1cm} (F_1+F_2+F_3)=0\ \ 
;\ \ -\sqrt\frac{3}{2}G_1=\eta\
\end{eqnarray}
The same relations can be derived respectively for $B_{bb}\to B_{bc}$
and $B_{bb}\to B_{bc}'$ decays.

\begin{figure}[t]
\resizebox{8cm}{!}{\includegraphics{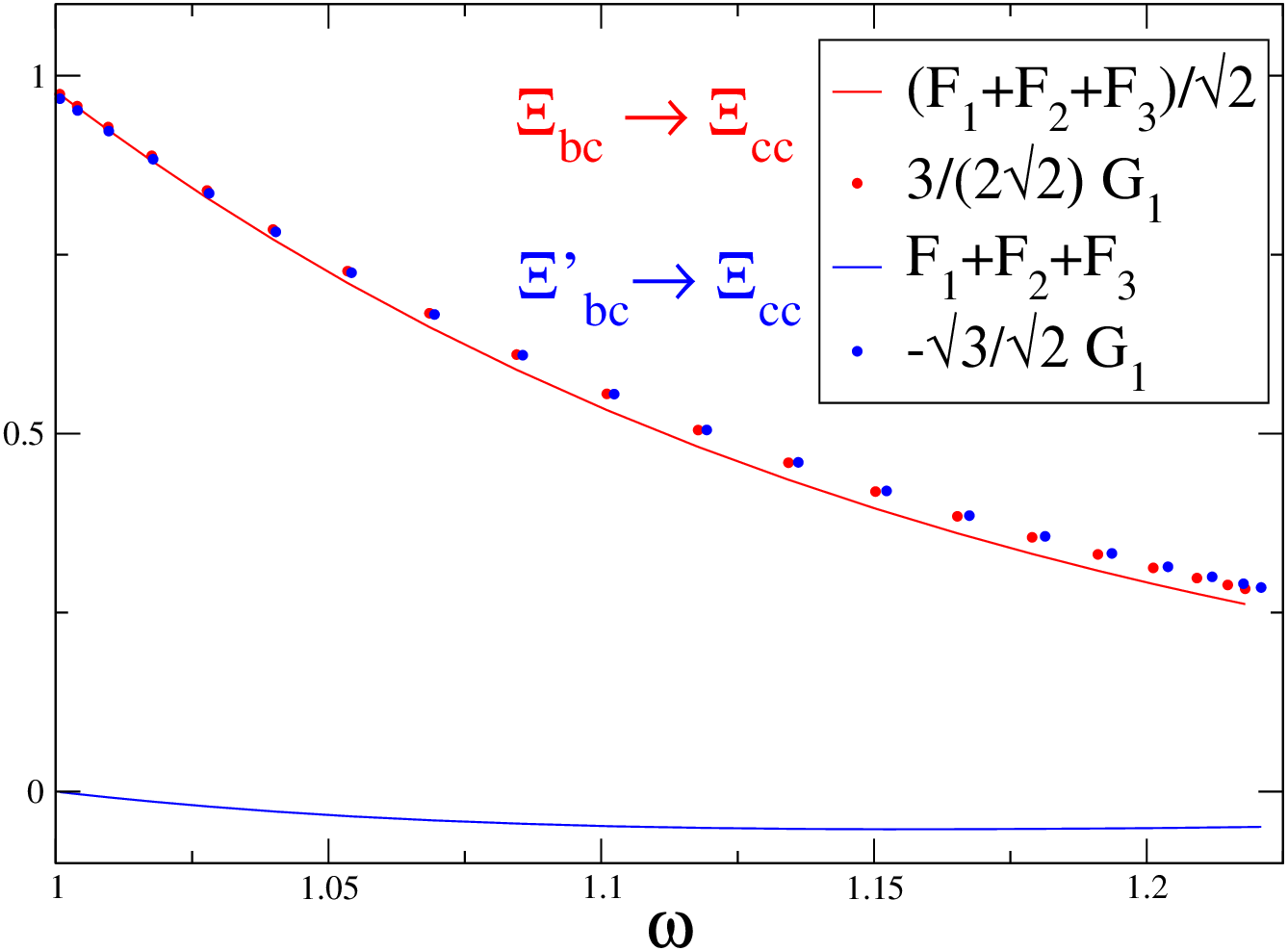}}
\hspace{1cm}\resizebox{8cm}{!}{\includegraphics{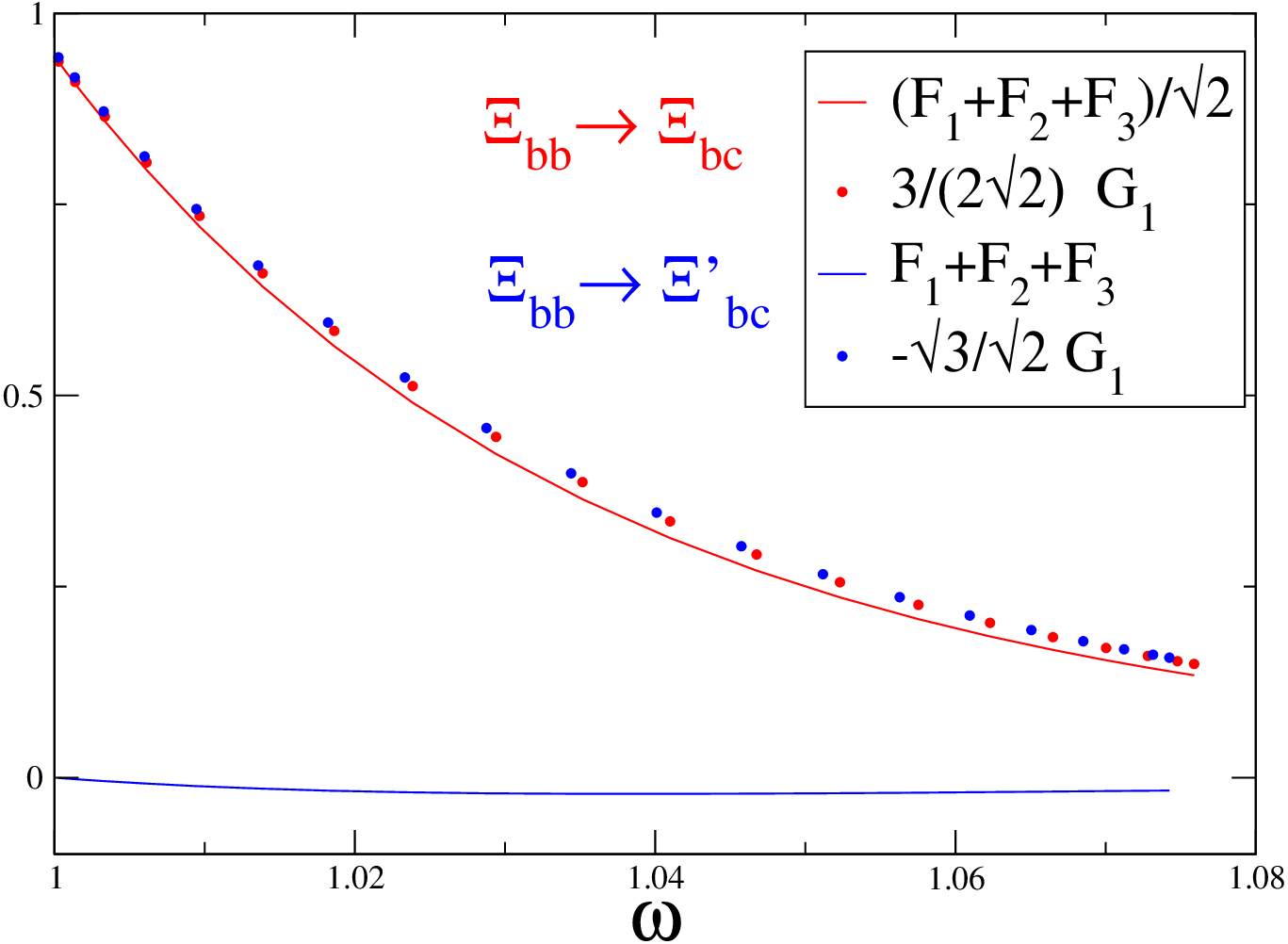}}
\vspace{.3cm}
\caption{ Left panel: $(F_1+F_2+F_3)/\sqrt2$ and $3\,G_1/2\sqrt2$ of
the $\Xi_{bc}\to\Xi_{cc}$ transition (in red), and $F_1+F_2+F_3$ and
$-\sqrt3\,G_1/\sqrt2$ of the $\Xi_{bc}'\to\Xi_{cc}$ transition (in
blue) evaluated using the AL1 interquark potential of
Ref.\cite{SS94}. Right panel: same as left panel for
$\Xi_{bb}\to\Xi_{bc}$ and $\Xi_{bb}\to\Xi_{bc}'$
transitions.}\vspace{.75cm}
\label{fig:1212}
\end{figure}

In Fig.\ref{fig:1212} we show our results for the above quantities
 evaluated for the $\Xi_{bc}\to\Xi_{cc}$, $\Xi_{bc}'\to\Xi_{cc}$ and
 $\Xi_{bb}\to\Xi_{bc}$, $\Xi_{bb}\to\Xi_{bc}'$ transitions. We have
 used the AL1 interquark potential of Ref.~\cite{SS94} and actual
 heavy quark masses (baryon masses are given in
 Table~\ref{tab:summ}). We do not show very similar results obtained
 for transitions involving $\Omega$ baryons. We see in the figure that
 the above restrictions are, to a good approximation, satisfied by our
 calculation over the entire $\omega$ region.

\subsection{$1/2\longleftrightarrow3/2$ transitions}
In this case one can see that all contributions generated by the
$C_3^V,\,C_4^V,\,C_5^V,\,C_6^V,\,$ and $C_6^A$ form factors are
proportional to $|\vec q\,|$, cancelling thus near zero recoil. The
same happens for the $q^\lambda\gamma^\mu$ dependence of the $C_3^A$
form factor and the $q^\lambda p'^\mu$ dependence of the $C_4^A$ form
factor. On the other hand the $g^{\lambda\mu}$ dependence of the axial
part of the current survives near zero recoil. The restrictions
imposed by Eqs.(\ref{eq:hqss3},\ref{eq:hqss4},\ref{eq:hqss5}) are in
this case

\begin{eqnarray}
\label{eq:ff1232_1}
&&B_{bc} \to B^*_{cc}\hspace{2cm}
-{\sqrt\frac32}\left(\frac{C_3^A}{m_{B_{bc}}}(m_{B_{b
c}}-m_{B^*_{cc}})
+\frac{C_4^A}{m^2_{B_{bc}}}(m_{B_{bc}}E_{B^*_{cc}}   -m_{B^*_{cc}}^2)+C_5^A\right)=\eta\\
\label{eq:ff1232_2}
&&B_{bc}' \to B^*_{cc}\hspace{2cm}
-\frac{1}{\sqrt2}\left(\frac{C_3^A}{m_{B'_{bc}}}(m_{B'_{bc}}-m_{B^*_{cc}})+\frac{C_4^A}{m^2_{B'_{bc}}}
(m_{B'_{bc}}E_{B^*_{cc}}-m_{B^*_{cc}}^2)+C_5^A\right)
=\eta\\
\label{eq:ff1232_3}
&&B^*_{bc} \to B_{cc}\hspace{2cm}
-{\sqrt\frac32}\left(-\frac{C_3^A}{m_{B_{cc}}}(m_{B^*_{bc}}-m_{B_{cc}})
-\frac{C_4^A}{m_{B_{cc}}^2}(m_{B^*_{bc}}^2-m_{B^*_{bc}}E_{B_{cc}})+C_5^A\right)=\eta
\end{eqnarray}

For $B_{bb}\to B_{bc}^*$, $B_{bb}^*\to B_{bc}$ and $B_{bb}^*\to
B_{bc}'$  the relations obtained are given respectively, and with obvious
changes,  by Eqs.(\ref{eq:ff1232_1}), (\ref{eq:ff1232_3})
and again (\ref{eq:ff1232_3}) but in the latter case with the factor
$\sqrt{3/2}$ changed to $1/\sqrt2$.

\begin{figure}[t]
\resizebox{8cm}{!}{\includegraphics{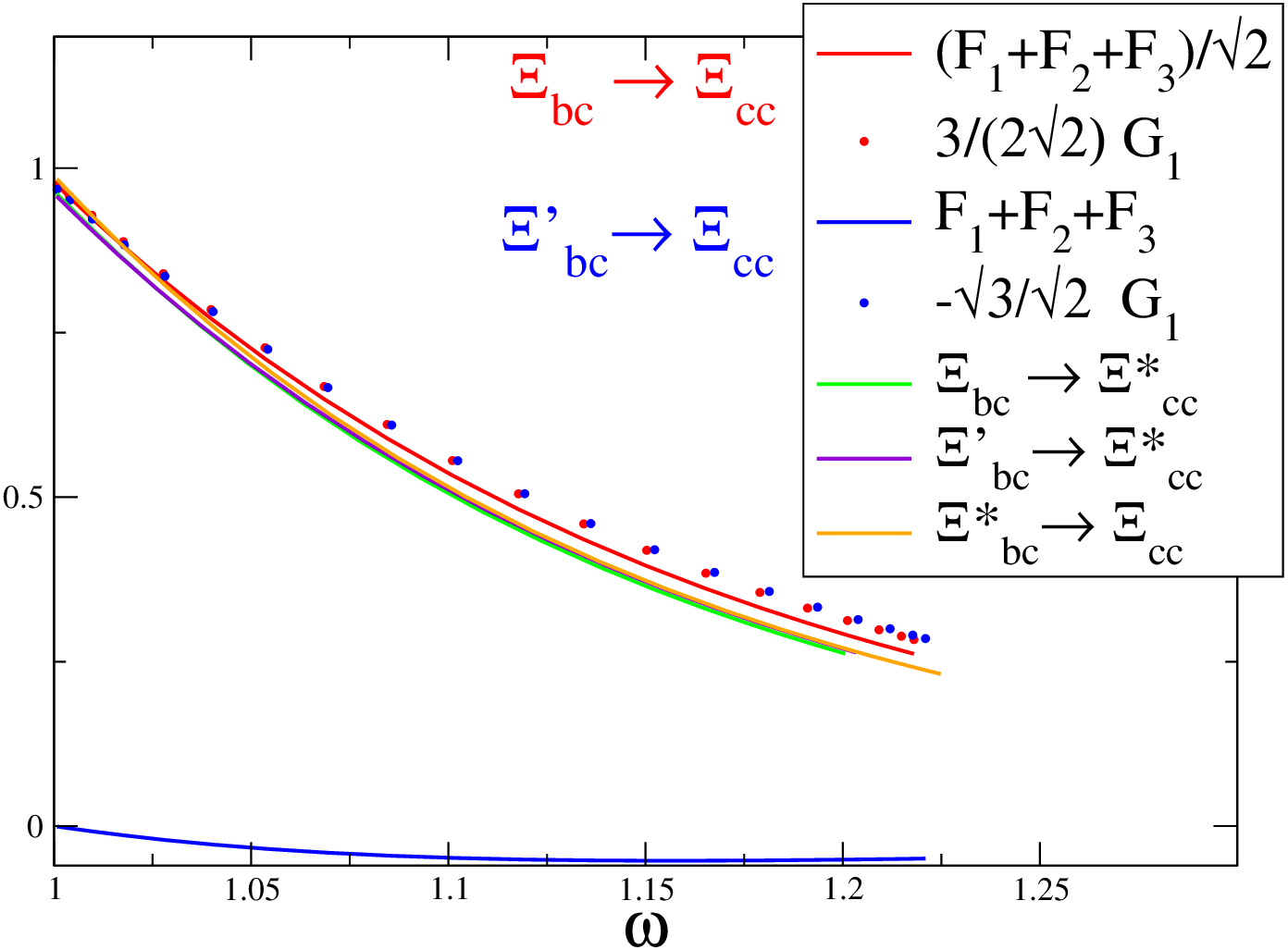}}\hspace{1cm}\resizebox{8cm}{!}{\includegraphics{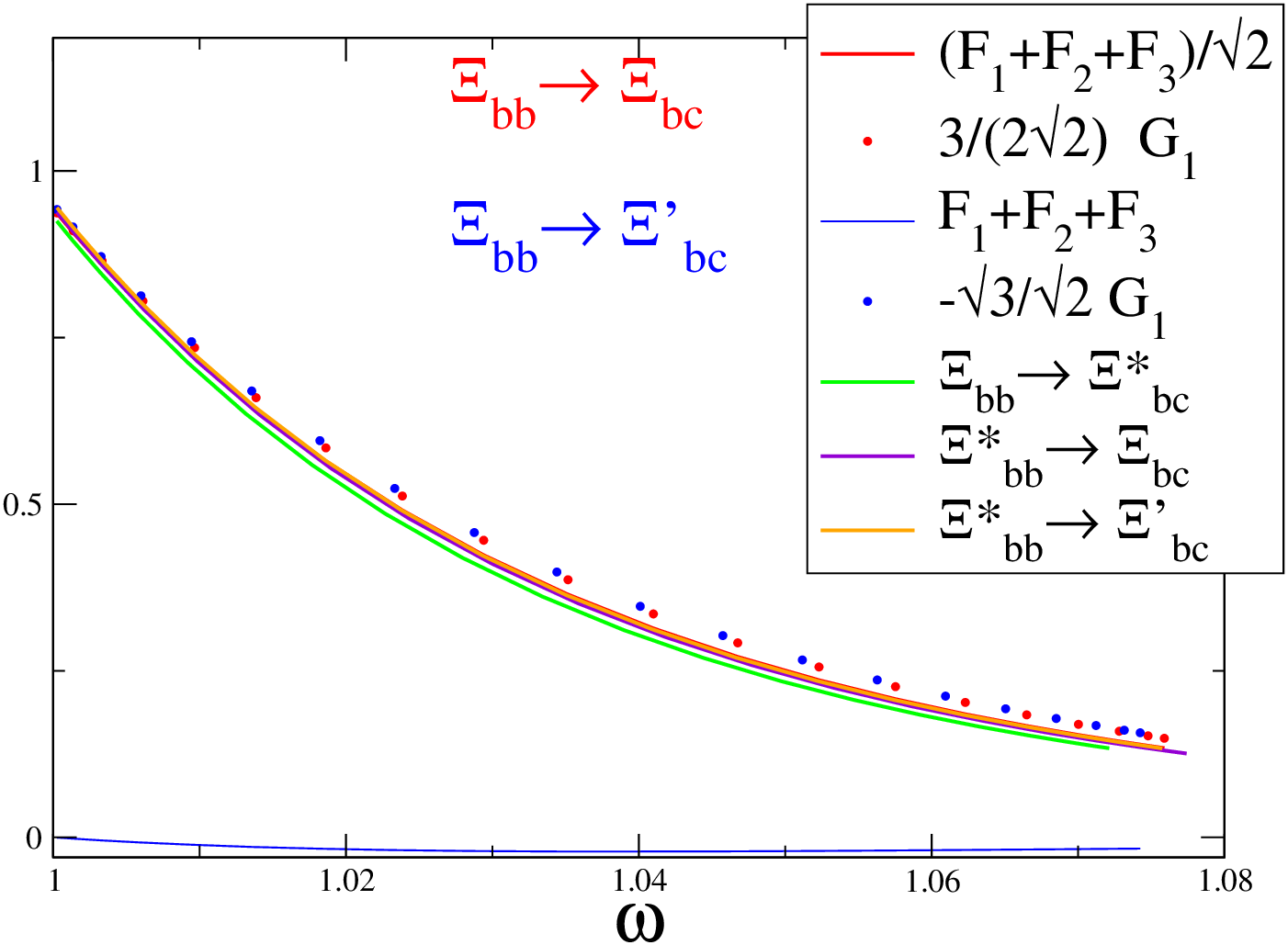}}
\vspace{.3cm}
\caption{Left panel: relations in
Eqs.~(\ref{eq:ff1232_1},\ref{eq:ff1232_2},\ref{eq:ff1232_3}) for
$\Xi_{bc}\to \Xi^*_{cc}$, $\Xi_{bc}'\to \Xi^*_{cc}$ and $\Xi^*_{bc}\to
\Xi_{cc}$ transitions. We also show, for better comparison the results
already shown in Fig.~\ref{fig:1212}.  Right panel: similar relations
for $\Xi_{bb}\to \Xi^*_{bc}$, $\Xi^*_{bb}\to \Xi_{bc}$ and
$\Xi^*_{bb}\to \Xi'_{bc}$ transitions. All the results have been
obtained using the AL1 interquark potential of Ref.\cite{SS94}.}
\label{fig:1232}
\end{figure}

We show now In Fig.\ref{fig:1232} our results for the
$\Xi_{bc}\to\Xi^*_{cc}$, $\Xi_{bc}'\to\Xi^*_{cc}$ and
$\Xi^*_{bc}\to\Xi^*_{cc}$ transitions, and for the
$\Xi_{bb}\to\Xi^*_{bc}$, $\Xi^*_{bb}\to\Xi_{bc}$ and
$\Xi^*_{bb}\to\Xi_{bc}'$ transitions, again evaluated with the AL1
interquark potential of Ref.\cite{SS94} and actual heavy quark
masses. We do not show results for $\Omega$ baryons which are very
similar to the ones presented. We see again our calculation is in
accordance with HQSS constraints.

\subsection{$3/2\to3/2$ transitions}
In this case we have not evaluated explicitly individual form factors.
Here we proceed as follows: we evaluate the hadronic matrix elements
in Eq.~(\ref{eq:hqss6}) for different  spin
configurations, selecting only vector or axial components for which
the matrix element does not cancel near zero recoil. The corresponding
matrix elements are also evaluated in the quark model of
Ref.~\cite{conrado07}. By comparison of the two calculations we get
the IW function. In this way one can obtain five different functions,
two of them with the vector part of the current and three others with
the axial part. In the infinite heavy quark mass limit these five
functions should coincide among themselves and with the ones obtained
in $1/2\to 1/2$ and $1/2\longleftrightarrow3/2$ transitions.

\begin{figure}[t]
\vspace{1cm}
\resizebox{8.5cm}{!}{\includegraphics{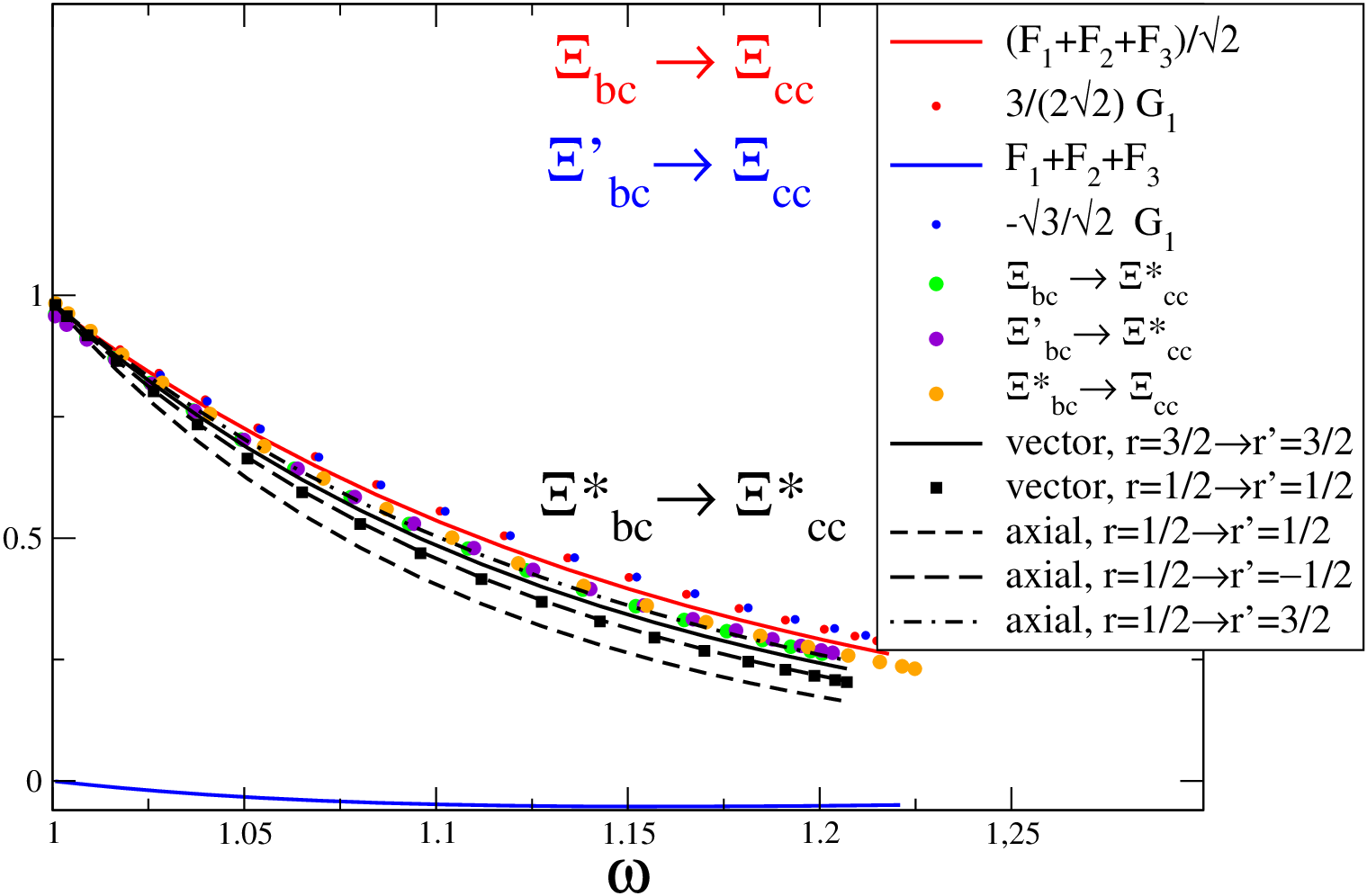}}\hspace{1cm}
\resizebox{8cm}{!}{\includegraphics{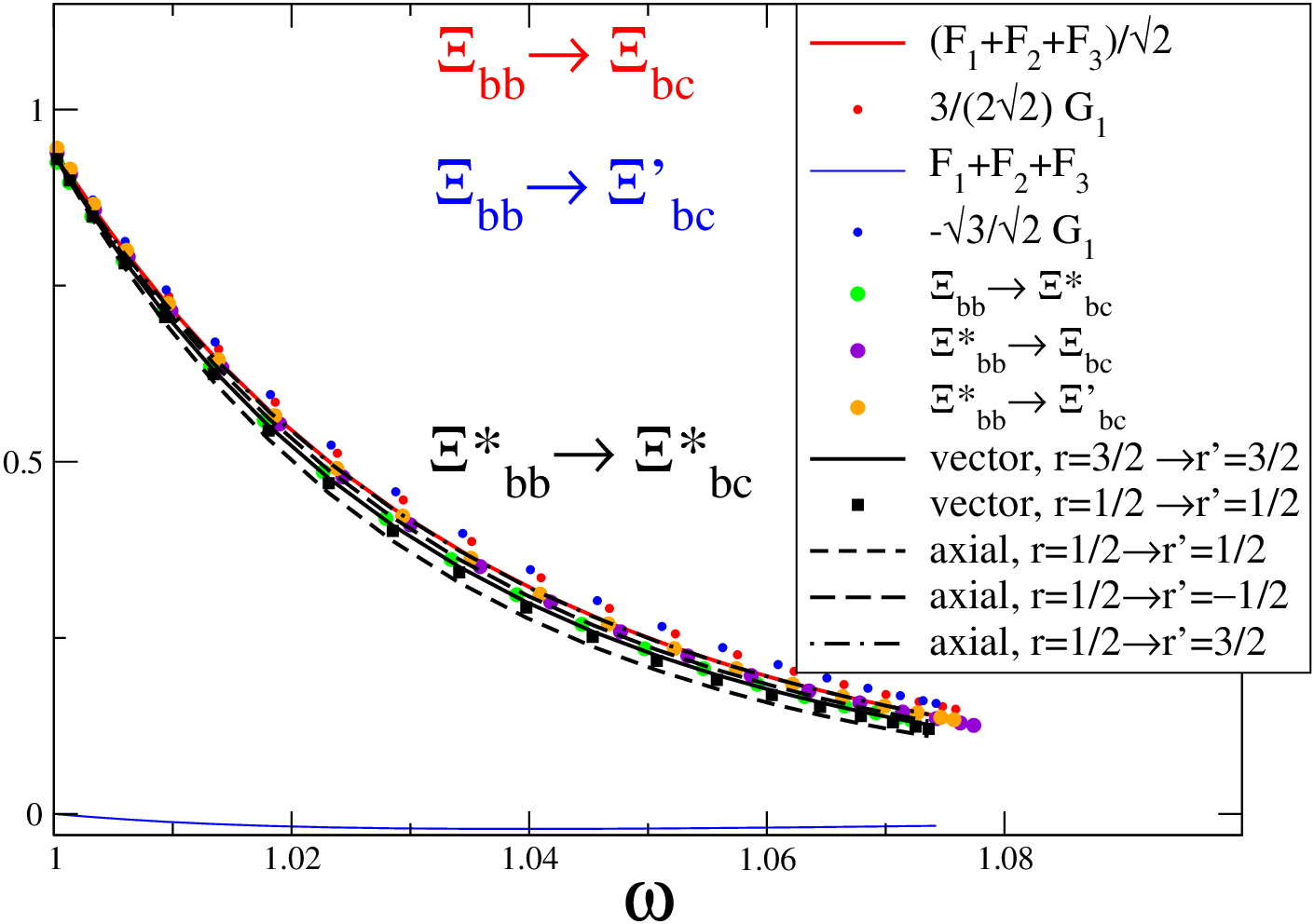}}
\vspace{0.3cm}
\caption{ Left panel: Different IW functions obtained for
$\Xi_{bc}^*\to\Xi_{cc}^*$ transitions (black curves) using the vector
or the axial part of the weak transition current, and for different
spin configurations. For better comparison we also show the
corresponding results obtained for $1/2\to1/2$ and
$1/2\longleftrightarrow3/2$ transitions.  All results have been
obtained using the AL1 interquark potential of Ref.\cite{SS94}. Right
panel: same as left panel for ${bb}\to{bc}$ transitions.}
\label{fig:3232}
\end{figure}

The results are shown in Fig.~\ref{fig:3232}. What we see is, that to
a good approximation, better in the $bb\to bc$ case as one is closer
to the infinite heavy quark mass limit, all $1/2\to1/2$,
$1/2\longleftrightarrow 3/2$ and $3/2\to3/2$ transitions are governed
in terms of just one function. As mention before this function is
different for the $bc\to cc$ and $bb\to bc$ cases due to HFS breaking.

\section{Semileptonic decay}
\label{sec:sd}
In this section we present our results for semileptonic decay widths
and compare them with the ones obtained in other quark model
approaches. In some cases there are large discrepancies between
different calculations. The fact that, at least for the $bb\to bc$
case, we are not far from the infinite heavy quark mass limit suggests
that some calculations might be inconsistent with HQSS.

The  decay width is given by
\begin{equation}
\Gamma= 
\frac{G_F^{\,2}}{2\pi^4}  |V_{cb}|^2{m_{B'}^3}
 \int_1^{\omega_{max}} d\omega\ \sqrt{\omega^2-1}\ 
{\cal L}^{\mu\nu}\,{\cal H}_{\mu\nu}
\end{equation}
where $G_F= 1.16637(1)\times10^{-11}$\,MeV$^{-2}$\cite{pdg06} is the
Fermi decay constant and $|V_{cb}|$ is the modulus of the
corresponding Cabibbo--Kobayashi--Maskawa matrix element. ${\cal
  L}^{\mu\nu}$ is the leptonic tensor defined as
\begin{eqnarray}
{\cal L}^{\mu\nu}&=& \int\frac{d^3k}{2E}\frac{d^3k'}{2E'}
\,\delta^{(4)}(q-k-k') \left(k'^\mu k^\nu +k'^\nu k^\mu
- g^{\mu\nu} k\cdot k^\prime + {\rm i}
\epsilon^{\mu\nu\alpha\beta}k'_{\alpha}k_\beta\right) \label{eq:lep}
\end{eqnarray}
where $k,\,k'$ represent the momenta of the final charged lepton and
antineutrino respectively.  We use the convention
$\epsilon^{0123}=-1$. Using Lorentz covariance one can write
\begin{eqnarray}
\label{eq:lt}
{\cal L}^{\mu\nu}=A(q^2)\,g^{\mu\nu}+
B(q^2)\,\frac{q^\mu q^\nu}{q^2}
\end{eqnarray}
where neglecting neutrino masses
\begin{eqnarray}
&&A(q^2)=-\frac{I(q^2)}{6}\left(2q^2-m_l^2-\frac{m_l^4}{q^2}\right)\nonumber\\
&&B(q^2)=\frac{I(q^2)}{3}\left({q^2+m_l^2}-2\frac{m_l^4}{q^2}\right)
\end{eqnarray}
with
\begin{eqnarray}
I(q^2)=\frac{\pi}{2q^2}(q^2-m_l^2)
\end{eqnarray}
Note that for a light lepton $l=e,\mu$ we can neglect terms in $m_l^2/q^2$ over most of the $q^2$ interval and thus use
$B(q^2)\approx-A(q^2)$.

The hadron tensor is given by
\begin{eqnarray}
{\cal H}_{\mu\nu}(p,p') = \frac{1}{2S+1} \sum_{r,r'}  
 &&\left\langle B'(S'), r'\
\vec{p}^{\,\prime}\left|\,
\overline \Psi^c(0)\gamma_\mu(I-\gamma_5)\Psi^b(0)\right| B(S), r\ \vec{p}   \right\rangle \nonumber\\
&&\hspace{-.35cm}\times \left\langle B'(S'), r'\ 
\vec{p}^{\,\prime}\left|\,\overline \Psi^c(0)\gamma_\nu(I-\gamma_5)
\Psi^b(0) \right|  B(S), r\ \vec{p} \right\rangle^*
\label{eq:wmunu}
\end{eqnarray}
In Ref.~\cite{conrado07} it is shown how $1/2\to1/2$ hadronic matrix
elements are evaluated within our model.
The extension to the $1/2\longleftrightarrow3/2$ and $3/2\to3/2$ cases 
is straightforward.\\

\begin{table}[h!!]
\begin{center}
\begin{tabular}{lcccc||lccccc
}
  &This work &\cite{ebert04}$^\dag$&\cite{guo98}&\cite{onishchenko00}& 
  &This work &\cite{ebert04}$^\dag$&\cite{guo98}&
  \cite{sanchis95}$^\ddag$&\cite{onishchenko00}\\ \hline
  &&&&&\\
\large$\Gamma(\Xi_{bb}\to\Xi_{bc}\,l\bar\nu_l)$\hspace{.5cm}  &$ 1.92^{+0.25}_{-0.05}$&
\hspace*{.15cm}1.63\hspace*{.15cm}&\hspace*{.15cm}28.5\hspace*{.15cm}&8.99&
\hspace*{.15cm}\large$\Gamma(\Xi_{bc}\to\Xi_{cc}\,l\bar\nu_l)$\hspace{.5cm} &$ 2.57^{+0.26}_{-0.03} $ & 2.30&
8.93&
8.0&8.87\\
&&&&&\\

\large$\Gamma(\Xi_{bb}\to\Xi_{bc}'\,l\bar\nu_l)$ &  $1.06 ^{+0.13}_{-0.03}$ &0.82&4.28&&
\hspace*{.15cm}\large$\Gamma(\Xi_{bc}'\to\Xi_{cc}\,l\bar\nu_l)$ &  $1.36^{+0.10}_{-0.03}$ & 0.88&7.76&\\
&&&&&\\
\large$\Gamma(\Xi_{bb}\to\Xi^*_{bc}\,l\bar\nu_l)$  &$ 0.61^{+0.04}$&
0.53&27.2&2.70&
\hspace*{.15cm}\large$\Gamma(\Xi_{bc}\to\Xi^*_{cc}\,l\bar\nu_l)$ &$0.75^{+0.06}$&0.72&14.1&
2.4&2.66\\ 
&&&&&\\
\large$\Gamma(\Xi^*_{bb}\to\Xi'_{bc}\,l\bar\nu_l)$  &$ 1.04^{+0.06} $ & 0.82&8.57&&
\hspace*{.15cm}\large$\Gamma(\Xi'_{bc}\to\Xi^*_{cc}\,l\bar\nu_l)$  &
$2.33^{+0.16}$& 1.70&28.8\\
&&&&& \\
\large$\Gamma(\Xi^*_{bb}\to\Xi_{bc}\,l\bar\nu_l)$ &  $0.35 ^{+0.03}$ &0.28&52.0&&
\hspace*{.15cm}\large$\Gamma(\Xi^*_{bc}\to\Xi_{cc}\,l\bar\nu_l)$ &$0.43^{+0.06}$& 0.38&27.5\\
&&&&& \\
\large$\Gamma(\Xi^*_{bb}\to\Xi^*_{bc}\,l\bar\nu_l)$ &  $2.09^{+0.16}$ &
1.92&12.9&&
\hspace*{.15cm}\large$\Gamma(\Xi^*_{bc}\to\Xi^*_{cc}\,l\bar\nu_l)$ &$2.63^{+0.40} $ &2.69&17.2\\ \hline
\end{tabular}
\caption{Decay widths in units of $10^{-14}$\,GeV for doubly heavy
$\Xi$ baryon semileptonic decay. Our central results have been
obtained with the AL1 potential of Ref.~\cite{SS94}. The errors show
the spread of results when using four other interquark potentials
taken from Refs.~\cite{SS94,BD81}.  We have used a value
\hbox{$|V_{cb}|=0.0413$}. $l$ stands for a light charged lepton,
$l=e,\,\mu$. For results with $^\dag$ and $^\ddag$ see text for
details. The results of Ref.~\cite{onishchenko00} are given as quoted
in Ref.~\cite{ebert04}. }
\label{tab:gammaxi}
\end{center}
\end{table}
In Table~\ref{tab:gammaxi} we compare our results for $\Xi\to\Xi$
transitions with the ones calculated in different models. Our central
values have been obtained with the AL1 potential of Ref.~\cite{SS94},
while the errors shown indicate the spread of the results when using
four other interquark potentials, three more taken from
Ref.~\cite{SS94} and another one from Ref.~\cite{BD81}. In all cases
we have used a value $|V_{cb}|=0.0413$.  Our results are in a global
reasonable agreement with the ones in Ref.~\cite{ebert04} where they
use a relativistic quark model evaluated in the quark-diquark
approximation\footnote{Note the results we show under
Ref.~\cite{ebert04} are a factor of 2 smaller than the originally
published. The reason beint that the authors of that reference also
missed a normalization factor $1/\sqrt2$ for diquarks with two equal
quarks~\cite{ebert08}. An erratum has been sent.}.  In
Ref.~\cite{sanchis95}, and using HQET, results around a factor of 4
larger than ours are found\footnote{Note the results we show under
Ref.~\cite{sanchis95} are a factor of 2 larger than the originally
published. There is a factor $\sqrt2$ wrong in the normalization of
matrix elements that affects the published
results~\cite{sanchis08}.}. We believe this factor of 4 discrepancy
stems from the fact that in Ref.~\cite{sanchis95} the author
approximates $\eta(\omega)$ by $\eta(1)$ in the calculation of the
decay width\footnote{If we take for instance the approximate
expression in the Eq.~(\ref{eq:sanchis}), which is closer to the
approximations used in Ref.~\cite{sanchis95}, and make
$\eta(\omega)=\eta(1)$ we get $\Gamma(\Xi_{bc}\to\Xi_{cc})=9.4\times
10^{-14}$\,GeV in agreement with the result in
Ref.~\cite{sanchis95}. On the other hand if we take the actual
$\eta(\omega)$ values we get $\Gamma(\Xi_{bc}\to\Xi_{cc})=2.4\times
10^{-14}$\,GeV, roughly a factor of 4 smaller and in agreement with
our full calculation result.}. In Ref.~\cite{onishchenko00} they
obtain results similar to the ones in the previous
reference\footnote{We must say we believe the calculation in
Ref.~\cite{onishchenko00} is affected by the same normalization
mistake made in the original calculation in Ref.~\cite{sanchis95} as
they give $F_1+F_2+F_3=\eta$ instead of $F_1+F_2+F_3=\sqrt2\,\eta$. To
our understanding their present results have to be multiplied by a
factor of 2.}.
Finally  in Ref.~\cite{guo98} they obtain in general
much larger results for all transitions. In this latter calculation
the authors take the Bethe--Salpeter equation model to analyze the
weak transition matrix elements between two heavy diquarks, and then
use ``superflavor''
symmetry~\cite{Georgi:1990ak,Savage:1990di,Carone:1990pv} to evaluate
the transition matrix elements at the baryon level.  The global
results show a contradiction between the calculation by Guo {\it et
al.}~\cite{guo98} in one hand, and ours and the one by Ebert {\it et
al.}~\cite{ebert04} on the other.  
%

%
%
%
%
\begin{table}[t]
\begin{center}
\begin{tabular}{lcc||lcc}
  &This work &\cite{ebert04}$^\dag$ 
 & &This work &\cite{ebert04}$^\dag$\\ \hline
  &&&&&\\
\large $\Gamma(\Omega_{bb}\to\Omega_{bc}\,l\bar\nu_l)$\hspace{.5cm}  &$
2.14^{+0.20}_{-0.02}$&\hspace*{.15cm}1.70\hspace*{.15cm}
&\hspace*{.15cm}\large $\Gamma(\Omega_{bc}\to\Omega_{cc}\,l\bar\nu_l)$\hspace{.5cm}  &$ 2.59^{+0.20} $
&\hspace*{.15cm}2.48\hspace*{.15cm} \\
&&&&&\\

\large $\Gamma(\Omega_{bb}\to\Omega_{bc}'\,l\bar\nu_l)$ &  $1.16 ^{+0.13}$ &0.83&
\large$\hspace*{.15cm} \Gamma(\Omega_{bc}'\to\Omega_{cc}\,l\bar\nu_l)$ &  $1.36^{+0.9}$ & 0.95\\
&&&&&\\
\large $\Gamma(\Omega_{bb}\to\Omega^*_{bc}\,l\bar\nu_l)$  &$ 0.67^{+0.08}$&
0.55&
\hspace*{.15cm}\large $\Gamma(\Omega_{bc}\to\Omega^*_{cc}\,l\bar\nu_l)$ &$0.76^{+0.13}$&0.74\\ 
&&&&&\\
\large $\Gamma(\Omega^*_{bb}\to\Omega'_{bc}\,l\bar\nu_l)$  &$ 1.13^{+0.11}_{-0.08} $ & 0.85&
\hspace*{.15cm}\large $\Gamma(\Omega'_{bc}\to\Omega^*_{cc}\,l\bar\nu_l)$  &
$2.36^{+0.33}$& 1.83\\
&&&&& \\
\large $\Gamma(\Omega^*_{bb}\to\Omega_{bc}\,l\bar\nu_l)$ &  $0.38 ^{+0.04}_{-0.02}$ &0.29&
\hspace*{.15cm}\large $\Gamma(\Omega^*_{bc}\to\Omega_{cc}\,l\bar\nu_l)$ &$0.44^{+0.06}$& 0.40\\
&&&&& \\
\large $\Gamma(\Omega^*_{bb}\to\Omega^*_{bc}\,l\bar\nu_l)$ &  $2.29^{+0.31}_{-0.04}$ &
2.0&
\hspace*{.15cm}\large $\Gamma(\Omega^*_{bc}\to\Omega^*_{cc}\,l\bar\nu_l)$ &$2.79^{+0.60} $ &2.88\\ \hline
\end{tabular}
\caption{Same as Table~\ref{tab:gammaxi} for doubly heavy $\Omega$
  baryon semileptonic decay. Decay widths are given in units of $10^{-14}$\,GeV.}
\label{tab:gammaomega}
\end{center}
\end{table}
In Table~\ref{tab:gammaomega} we show results for $\Omega\to\Omega$
transitions. Again we get a global reasonable agreement with the
calculation by Ebert {\it et al.}~\cite{ebert04}. %

It is worthwhile to mention that for the case of baryons with a $bc$ heavy quark content the actual physical states
$\Xi$ and $\Omega$
 will be an admixture of  $\Xi_{bc},\,\Xi'_{bc}$ and $\Omega_{bc},\,\Omega'_{bc}$ respectively. If we look for instance at our
 model predictions we see the widths are very different for transitions involving $\Xi_{bc}$ or $\Xi'_{bc}$,
 and $\Omega_{bc}$ or $\Omega'_{bc}$. Accurate measurements of decay widths could thus give
 information on the admixtures.%
\section{HQSS constraints on semileptonic decay widths}
\label{sect:hqsswidth}
To the extent that one is close enough to the infinite heavy quark
mass limit and near zero recoil we can combine the HQSS results in
Eqs.(\ref{eq:hqss1}-\ref{eq:hqss6}) with Eq.(\ref{eq:lt}), to
approximate the tensor product ${\cal L}^{\mu\nu}\,{\cal H}_{\mu\nu}$
by
\begin{eqnarray}
B_{bc}\to B_{cc}\hspace{1.cm}{\cal L}^{\mu\nu}\,{\cal H}_{\mu\nu}&\approx&\eta^2\frac{1}{9}\bigg\{
A(q^2)\left(-26\, \omega+20\right)
\left.+B(q^2)\left[26\,\frac{(v'\cdot q)(v\cdot
q)}{q^2}+(5-13\,\omega)\right]\right\}\label{eq:sanchis}\\
B'_{bc}\to B_{cc}\hspace{1cm}{\cal L}^{\mu\nu}\,{\cal H}_{\mu\nu}&\approx&\eta^2\frac{1}{9}\bigg\{
A(q^2)\left(-6\, \omega-12\right)\left.+B(q^2)\left[6\,\frac{(v'\cdot q)(v\cdot
q)}{q^2}-3(1+\,\omega)\right]\right\}\\
B_{bc}\to B^*_{cc}\hspace{1cm}{\cal L}^{\mu\nu}\,{\cal H}_{\mu\nu}&\approx&\eta^2
\frac{1+\omega}{9}\left\{
-6A(q^2)+2B(q^2)\left[\,\frac{(v'\cdot q)^2}{q^2}-1\right]\right\}\\
B'_{bc}\to B^*_{cc}\hspace{1cm}{\cal L}^{\mu\nu}\,{\cal H}_{\mu\nu}&\approx&\eta^2
\frac{1+\omega}{3}\left\{
-6A(q^2)+2B(q^2)\left[\,\frac{(v'\cdot q)^2}{q^2}-1\right]\right\}\\
B^*_{bc}\to B_{cc}\hspace{1cm}{\cal L}^{\mu\nu}\,{\cal H}_{\mu\nu}&\approx&\eta^2
\frac{1+\omega}{9}\left\{
-3A(q^2)+B(q^2)\left[\,\frac{(v\cdot q)^2}{q^2}-1\right]\right\}\\
B^*_{bc}\to B^*_{cc}\hspace{1cm}{\cal L}^{\mu\nu}\,{\cal H}_{\mu\nu}&\approx&\eta^2\frac{1}{9}\left\{
-A(q^2)\,\omega\left(4+8\omega^2
\right)
%
%
%
+B(q^2)\left[
-\,\omega\left(6+4\omega^2
\right)
+\frac{(v'\cdot q)(v\cdot q)}{q^2}
\left(20+8\omega^2
\right)\right]
\right\}\nonumber\\
\end{eqnarray}
and similar ones for $bb\to bc $ decays. 

Working in the strict near zero recoil approximation, $\omega\approx
1$ or equivalently $q^2$ quite close to its maximum value $q^2_{\rm max}$,
we can approximate
\begin{equation}
\frac{(v\cdot q)^2}{q^2}\approx\frac{(v'\cdot q)(v\cdot
  q)}{q^2}\approx\frac{(v'\cdot q)^2}{q^2} \approx 1,
\end{equation}
and $A(q^2)\approx -B(q^2)$ near $q^2_{\rm max}$. In these
circumstances, and using
\begin{eqnarray}
m_{B_{bb}}\approx m_{B^*_{bb}}\ ;\ m_{B_{bc}}\approx m_{B'_{bc}}\approx m_{B^*_{bc}}\ ;\
m_{B_{cc}}\approx m_{B^*_{cc}},
\end{eqnarray}
HQSS predicts that the different decay widths are in the relative ratios
\begin{eqnarray}
\label{eq:snzrbc}
&&\Gamma(B_{bc}\to B_{cc})\ :\ \Gamma(B'_{bc}\to B_{cc})\ :\ \Gamma(B_{bc}\to B^*_{cc})\ :
\ \Gamma(B'_{bc}\to B^*_{cc})\ :\ \Gamma(B^*_{bc}\to B_{cc})\ :
\ \Gamma(B^*_{bc}\to B^*_{cc})
\nonumber\\
&&\hspace{1cm}4\hspace{1.05cm}:\hspace{1.1cm}3\hspace{1.05cm}:\hspace{1.15cm}2\hspace{1.05cm}:
\hspace{1.1cm}6\hspace{1.05cm}:\hspace{1.15cm}1\hspace{1.05cm}:\hspace{1.1cm}5\hspace{1cm}
\end{eqnarray}
\begin{eqnarray}
\label{eq:snzrbb}
&&\Gamma(B_{bb}\to B_{bc})\ :\ \Gamma(B_{bb}\to B'_{bc})\ :\ \Gamma(B_{bb}\to B^*_{bc})\ :
\ \Gamma(B^*_{bb}\to B'_{bc})\ :\ \Gamma(B^*_{bb}\to B_{bc})\ :
\ \Gamma(B^*_{bc}\to B^*_{bc})
\nonumber\\
&&\hspace{1cm}4\hspace{1.05cm}:\hspace{1.1cm}3\hspace{1.05cm}:\hspace{1.15cm}2\hspace{1.05cm}:
\hspace{1.1cm}3\hspace{1.05cm}:\hspace{1.15cm}1\hspace{1.05cm}:\hspace{1.1cm}5\hspace{1cm}
\end{eqnarray}
In Table~\ref{tab:hqsr} we show the above ratios obtained in different
models. Our results and the ones by Ebert {\it et al.}\cite{ebert04}
are in reasonable agreement with the HQSS predictions in this strict
near zero recoil approximation.  On the other hand the results by Guo
{\it et al.}\cite{guo98} deviate heavily form the above
predictions. This disagreement does not improve much by using a
different decay width to normalize the ratios

\begin{table}[h!!]
\begin{tabular}{ll|cccccccccccc}
&&$\Gamma(B_{bc}\to B_{cc})$&:&$\Gamma(B'_{bc}\to B_{cc})$&:&$\Gamma(B_{bc}\to B^*_{cc})$&:&
$\Gamma(B'_{bc}\to B^*_{cc})$&:&$\Gamma(B^*_{bc}\to B_{cc})$&:&
$\Gamma(B^*_{bc}\to B^*_{cc})$\\\hline
&&&&&\\
HQSS&&4&:&3&:&2&:&6&:&1&:&5\\
This work&$\Xi$&6.04&:&3.20&:&1.75&:&5.48&:&1&:&6.18\\
This work&$\Omega$&5.88&:&3.08&:&1.73&:&5.36&:&1&:&6.33\\
\cite{ebert04}&$\Xi$&6.12&:&2.35&:&1.91&:&4.53&:&1&:&7.16\\
\cite{ebert04}&$\Omega$&6.19&:&2.38&:&1.85&:&4.58&:&1&:&7.20\\
\cite{guo98}&$\Xi$&0.32&:&0.28&:&0.53&:&1.05&:&1&:&0.63\\\\
&&&\\
&&$\Gamma(B_{bb}\to B_{bc})$&:&$\Gamma(B_{bb}\to B'_{bc})$&:&$\Gamma(B_{bb}\to B^*_{bc})$&:&
$\Gamma(B^*_{bb}\to B'_{bc})$&:&$\Gamma(B^*_{bb}\to B_{bc})$&:&
$\Gamma(B^*_{bb}\to B^*_{bc})$\\\hline
&&&&&\\
HQSS&&4&:&3&:&2&:&3&:&1&:&5\\
This work&$\Xi$&5.56&:&3.07&:&1.75&:&3.01&:&1&:&6.04\\
This work&$\Omega$&5.71&:&3.09&:&1.77&:&3.01&:&1&:&6.11\\
\cite{ebert04}&$\Xi$&5.93&:&2.98&:&1.91&:&2.96&:&1&:&6.96\\
\cite{ebert04}&$\Omega$&5.96&:&2.91&:&1.93&:&2.98&:&1&:&7.00\\
\cite{guo98}&$\Xi$&0.55&:&0.08&:&0.52&:&0.16&:&1&:&0.24\\\\
\end{tabular}
\caption{Decay width ratios for semileptonic $bc\to cc$ and $bb\to bc
  $ decay of doubly heavy $\Xi$ and $\Omega$ baryons compared to the HQSS predictions  in the 
  strict near zero recoil approximation.  Our results have
  been obtained with the AL1 potential of Ref.~\cite{SS94}.  $l$
  stands for a light charged lepton, $l=e,\,\mu$.}
\label{tab:hqsr}
\end{table}

We can relax the strict near near zero recoil approximation to obtain
more accurate predictions based on HQSS in the following way. For the
actual doubly heavy baryon masses $\omega_{max}\approx 1.22\ (1.08)$
for $bc\to cc\ (bb\to bc)$ transitions while the different
differential decay widths $d\Gamma/d\omega$ show a maximum at around
$\omega\approx 1.05\ (1.01)$. We can thus still use $\omega\approx1$
and $A(q^2)\approx -B(q^2)$. On the other hand the quantities $(v\cdot
q)^2/q^2,\,(v'\cdot q)^2/q^2,\,(v\cdot q)(v'\cdot q)/q^2$, that are
all equal to 1 near zero recoil, can deviate rapidly from 1 because of
the $q^2$ factor in the denominator.  What is true, in and around the
maximum of the differential decay width, is that we can reasonable
approximate
\begin{eqnarray}
\frac{(v\cdot q)^2}{q^2}\approx\frac{(v'\cdot q)(v\cdot q)}{q^2}\nonumber\\
\frac{(v'\cdot q)^2}{q^2}\approx\frac{(v'\cdot q)(v\cdot q)}{q^2}
\end{eqnarray}
With the above consideration we can still predict  approximate ratios
between different decay widths that one expects to be satisfied to an
accuracy of  $20\sim30$\%. We have chosen to define those
ratios so that they are all equal to one,
\begin{eqnarray}
 \label{eq:hqssdw1}
 \frac{\Gamma(B'_{bc}\to  B^*_{cc}\,l\bar\nu_l)}
{3\,\Gamma( B_{bc}\to  B^*_{cc}\,l\bar\nu_l)}&\approx& 
 \frac{\Gamma(B^*_{bb}\to  B'_{bc}\,l\bar\nu_l)}
{3\,\Gamma( B^*_{bb}\to  B_{bc}\,l\bar\nu_l)}\approx1 \\
 \label{eq:hqssdw2}
\frac{\Gamma({ B}_{bc}\to { B}^*_{cc}\,l\bar\nu_l)}{\frac23\,\Gamma( B'_{bc}\to
{ B}_{cc}\,l\bar\nu_l) }&\approx& 
\frac{\Gamma({ B}_{bb}\to { B}^*_{bc}\,l\bar\nu_l)}{\frac23\,\Gamma({ B}_{bb}\to
 B'_{bc}\,l\bar\nu_l) }\approx 1\\
 \label{eq:hqssdw3}
\frac{\Gamma({ B}^*_{bc}\to { B}_{cc}\,l\bar\nu_l)}{\frac13\,\Gamma(B'_{bc}\to{ B}_{cc}\,l\bar\nu_l)}&\approx&
\frac{\Gamma({ B}^*_{bb}\to { B}_{bc}\,l\bar\nu_l)}
{\frac13\,\Gamma({ B}_{bb}\to B'_{bc}\,l\bar\nu_l)}\approx1\\
 \label{eq:hqssdw4}
\frac{\Gamma({ B}^*_{bc}\to { B}^*_{cc}\,l\bar\nu_l)}{\Gamma({ B}_{bc}\to
{ B}_{cc}\,l\bar\nu_l)+\frac12\,\Gamma({ B}_{bc}\to 
{ B}^*_{cc}\,l\bar\nu_l)}&\approx&
\frac{\Gamma({ B}^*_{bb}\to { B}^*_{bc}\,l\bar\nu_l)}{\Gamma({ B}_{bb}\to
{ B}_{bc}\,l\bar\nu_l)+\frac12\,\Gamma({ B}_{bb}\to 
{ B}^*_{bc}\,l\bar\nu_l)}\approx1
\end{eqnarray}                        
Note, we consider as independent the
phase-space integrals of $\eta^2(\omega)A(q^2)$ and
$\eta^2(\omega)A(q^2)\frac{(v'\cdot q)(v\cdot q)}{q^2}$. 

In Tables~\ref{tab:dwxi},\ref{tab:dwomega} we show the above ratios
evaluated in different quark model approaches. Once again calculations in this
work and the ones in Ref.\cite{ebert04} are compatible, within the
expected accuracy, with the approximate ratios obtained using HQSS
results. On the other hand the deviations found in the results by Guo
{\it et al.}  are, in most cases, too large.

\begin{table}[h!!]
\begin{tabular}{c|cccccc}
\multicolumn{1}{c|}{}&\multicolumn{2}{c}{This work}&\multicolumn{2}{c}{\protect{\cite{ebert04}}}
&\multicolumn{2}{c}{\protect{\cite{guo98}}}\\
&$\Xi$ &$\Omega$ &$\Xi$ &$\Omega$ &$\Xi$ &$\Omega$ \\\hline
&&&&&\\
\Large $\frac{\Gamma(B'_{bc}\to  B^*_{cc}\,l\bar\nu_l)}{3\,\Gamma( B_{bc}\to  B^*_{cc}\,l\bar\nu_l)}$&$1.04^{+0.03}_{-0.01}$&
$1.04_{-0.03}$&0.79&0.82&0.68&---\\
 &&&&&\\
\Large $\frac{\Gamma({ B}_{bc}\to { B}^*_{cc}\,l\bar\nu_l)}{\frac23\,\Gamma({ B}'_{bc}\to
{ B}_{cc}\,l\bar\nu_l) }$&$0.82^{+0.06}_{-0.01}$&$0.84^{+0.13}_{-0.01}$&1.22&1.17&2.72&---\\
 &&&&&\\
\Large $\frac{\Gamma({ B}^*_{bc}\to { B}_{cc}\,l\bar\nu_l)}{\frac13\,\Gamma(B'_{bc}\to{ B}_{cc}\,l\bar\nu_l)}$&$0.94^{+0.11}$&
$0.97^{+0.10}_{-0.01}$&1.28&1.26&10.6&---\\
 &&&&&\\
\Large $\frac{\Gamma({ B}^*_{bc}\to { B}^*_{cc}\,l\bar\nu_l)}{\Gamma({ B}_{bc}\to
{ B}_{cc}\,l\bar\nu_l)+\frac12\,\Gamma({ B}_{bc}\to { B}^*_{cc}\,l\bar\nu_l)}$&$0.89^{+0.11}$&$0.94^{+0.13}_{-0.01}$&1.01&1.01&1.08&---\\\hline
\end{tabular}
\caption{Decay width ratios for 
  semileptonic $bc\to cc$ decay of doubly heavy $\Xi$ and $\Omega$
  baryons.  In all cases the  approximate result 
  obtained using HQSS is 1. Our central results  
  have been obtained  with the AL1 potential of Ref.~\cite{SS94}. The
  errors show the spread of results when using four other 
  interquark potentials taken from  Refs.~\cite{SS94,BD81}.  
  $l$ stands for a light charged lepton, $l=e,\,\mu$.}
\label{tab:dwxi}
\end{table}
\begin{table}[h!]
\begin{tabular}{c|cccccc}
\multicolumn{1}{c|}{}&\multicolumn{2}{c}{This work}&\multicolumn{2}{c}{\protect{\cite{ebert04}}}
&\multicolumn{2}{c}{\protect{\cite{guo98}}}\\
&$\Xi$ &$\Omega$ &$\Xi$ &$\Omega$ &$\Xi$ &$\Omega$ \\\hline
&&&&&\\
\Large $\frac{\Gamma(B^*_{bb}\to  B'_{bc}\,l\bar\nu_l)}{3\,\Gamma( B^*_{bb}\to  B_{bc}\,l\bar\nu_l)}$&$1.00^{+0.01}_{-0.04}$&
$1.00^{+0.03}_{-0.01}$&0.99&0.99&0.05&---\\
 &&&&&\\
\Large $\frac{\Gamma({ B}_{bb}\to { B}^*_{bc}\,l\bar\nu_l)}{\frac23\,\Gamma({ B}_{bb}\to
 B'_{bc}\,l\bar\nu_l) }$&$0.86^{+0.08}_{-0.06}$&$0.86^{+0.05}$&0.96&0.99&9.53&---\\
 &&&&&\\
\Large $\frac{\Gamma({ B}^*_{bb}\to { B}_{bc}\,l\bar\nu_l)}{\frac13\,\Gamma({ B}_{bb}\to
B'_{bc}\,l\bar\nu_l)}$&$0.98^{+0.09}_{-0.03}$&
$0.97^{+0.06}_{-0.14}$&1.01&1.03&36.4&---\\
 &&&&&\\
\Large $\frac{\Gamma({ B}^*_{bb}\to { B}^*_{bc}\,l\bar\nu_l)}{\Gamma({ B}_{bb}\to
{ B}_{bc}\,l\bar\nu_l)+\frac12\,\Gamma({ B}_{bb}\to { B}^*_{bc}\,l\bar\nu_l)}$&$0.94^{+0.07}_{-0.06}$&$0.93^{+0.11}_{-0.10}$&1.01&1.01&0.31&---\\\hline
\end{tabular}
\caption{Same as Table~\ref{tab:dwxi} for semileptonic $bb\to bc$ decay of doubly heavy $\Xi$ and $\Omega$ baryons.}
\label{tab:dwomega}
\end{table}

\section{Summary}
\label{sec:summary}
We have checked the constraints imposed by HQSS on form factors and
decay widths. To our knowledge those constraints have not been
exploited before to check the consistency of different quark model
calculations.  We have shown that our calculation is consistent with
HQSS. The ratios in Eqs.(\ref{eq:snzrbc},\ref{eq:snzrbb}), obtained
using HQSS with strict zero recoil approximation, and the approximate
ratios in Eqs.(\ref{eq:hqssdw1}-\ref{eq:hqssdw4}), where we have
relaxed that approximation, compare well with the results in our model
and the one by Ebert {\it et al.}~\cite{ebert04}, but they are
incompatible with the calculation in Ref.~\cite{guo98}. We think that
although this is not enough guarantee for the predictions here and in
Ref.~\cite{ebert04} to be fully correct (in fact the few results in
Refs.~\cite{sanchis95,onishchenko00} are not incompatible with HQSS constraints
while they are a factor of four larger than ours), it certainly
indicates problems either in the model or in the calculation performed
in Ref.~\cite{guo98}.

\begin{acknowledgments}
 This research was supported by DGI and FEDER funds, under contracts
FIS2005-00810, FIS2006-03438, FPA2007-65748, and the Spanish
Consolider-Ingenio 2010 Programme CPAN (CSD2007-00042), by Junta de
Andaluc\'\i a and Junta de Castilla y Le\'on under contracts FQM0225
and SA016A07, and it is part of the EU integrated infrastructure
initiative Hadron Physics Project under contract number
RII3-CT-2004-506078.
\end{acknowledgments}

\end{document}